\documentclass[11pt]{article}
\usepackage{epsfig}
\usepackage{amstex}

\numberwithin{equation}{section}

\title{An analytic description of the vector constrained KP hierarchy}
\author{G.F. Helminck\hskip 2.cm  J.W. van de Leur\footnote{JvdL
is financially supported by the
Netherlands Organization for Scientific Research (NWO).}\\
\\
\\
Faculty of Applied Mathematics,\\
University of Twente,\\
P.O.Box 217, 7500 AE  Enschede,\\
The Netherlands\\
fax: 31-53 489 4824\\
e-mail: helminck@@math.utwente.nl\\
\hskip .9cm vdleur@@math.utwente.nl\\
}

\newtheorem{corollary}{Corollary}[section]

\newtheorem{lemma}{Lemma}[section]
\newtheorem{proposition}{Proposition}[section]
\newtheorem{theorem}{Theorem}[section]

\begin{document}

\maketitle

\begin{tabular}[h]{lcl}
1991 MSC &:& 22E65, 22E70, 35Q53, 35Q58, 58B25\\
Keywords &:& KP hierarchy, Constrained KP, Grassmann manifold
\end{tabular}

\begin{abstract}
  In this paper we give a geometric description in terms of the
  Grassmann manifold of Segal and Wilson, of the reduction of the KP
  hierarchy known as the vector $k$-constrained KP hierarchy. We also
  show in a geometric way that these hierarchies are equivalent to
  Krichever's general rational reductions of the KP hierarchy.
\end{abstract}

\section{Introduction}

In recent years (vector) constrained KP hierarchies have attracted
considerable attention both from the mathematical as the physical
community \cite{A}-\cite{Or1}, \cite{SS}, \cite{Z}, \cite{ZC}.
Many interesting
integrable systems like the
AKNS, Yajima--Oikawa and Melnikov
hierarchies appear amongst these
constrained  families. In the physics literature they are studied
in connection with multi-matrix models.

The (vector) constrained KP hierarchies were introduced as  reductions of the
KP
hierarchy
\[
  \frac{\partial L}{\partial t_n} =
  [(L^n)_+ , L],\quad n\geq 1,
\] for the first order pseudodifferential operator
$L=\partial + \sum_{j<0} \ell_j \partial^j$.
This reduction consists of assuming that
\[
(L^k)_-=\sum_{j=1}^m q_j\partial^{-1}r_j,
\]
such that the following conditions on the functions $q_j$ and $r_j$
hold:
\[
\frac {\partial q_j}{\partial t_n} = (L^n_W)_+ (q_j) \quad\mbox{and}\quad
 \frac {\partial r_j}{\partial t_n}= -(L^n_W)^*_+ (r_j) \quad \mbox{{}for all}
\;\;n\geq 1.
\]
In this way it generalizes the well-known Gelfand-Dickey hierarchies
($(L^k)_-=0$).

Much is known about these constrained hierarchies
and many well-known features are investigated,
e.g. it was shown that they possess a bi-Hamiltonian structure {\cite
{C}, \cite {Li}, \cite{OS},  \cite {SS}, \cite {ZC}, a
bilinear representation \cite{CZ}, \cite{LW}, \cite{LW1}, \cite{ZC}
and B\"acklund-Darboux and Miura
transformations \cite{A}, \cite{AGZ}, \cite{ANP}, \cite{ANP2}, \cite{ANP3},
\cite{C1}, \cite{MR}.
However, until recently, the geometry remained unclear.
It is well-known that  one can
associates to a point in an
infinite Grassmannian  a solution $L$ of the KP hierarchy \cite{S},
\cite{SW}.
In this paper we consider the
Segal-Wilson Grassmannian.
Let $H$ be the Hilbert space
of all square integrable functions on the circle
$S^1=\{ z\in\mathbb{C}\mid |z|=1\}$, which decomposes in
a natural way as the direct sum of two infinite dimensional
orthogonal closed
subspaces $H_+=\{\sum_{n\ge 0} a_nz^n\in H\}$ and $H_-=\{\sum_{n< 0}
a_nz^n\in H\}$.
The Segal-Wilson Grassmannian $Gr(H)$
consists of all closed subspaces $W\subset H$ such that the orthogonal
projection on
 $H_-$ is a Hilbert-Schmidt operator.
In this setting, the $k$-th Gelfand-Dickey hierarchy has the following simple
geometrical interpretation. The KP operator  $L$ belongs to the $k$-th
Gelfand-Dickey
hierarchy if and only if the corresponding $W\in Gr(H)$ satisfies
$z^kW\subset W$.
One of the authors gave in \cite{L1} (see also  \cite {L}) a
simple interpretation of the constrained KP hierarchy
for the case of polynomial tau-functions, viz
$L$ belongs to the $m$-vector $k$-constrained KP hierarchy
if and only if the corresponding $W\in Gr(H)$ has a subspace $W'$ of
codimension $m$ such that $z^k(W')\subset W$. We show
in this paper
that the  same interpretation also holds in the Segal-Wilson case.
Using this geometrical interpretation,  we prove in section 5 that the
vector constrained KP hierarchy describes the same reduction of KP as
the general rational reductions of Krichever
\cite{K} (see also \cite{D1}).
Our geometrical interpretation is  also  useful to give
solutions of these hierarchies (see e.g. \cite{L1}).

\section{The KP hierarchy revisited}

In this section we recall some results {}for the KP-hierarchy that we
will need in this paper. The KP hierarchy starts with a commutative
ring $R$ and a privileged derivation $\partial$ of $R$. In order to be
able to take roots of differential operators in $\partial$ with
coefficients {}form $R$, one extends this ring $R[\partial ]$ to the
ring $R[\partial , \partial^{-1})$ of pseudodifferential operators
with coefficients in $R$. It consists of all expressions
\[
\sum^N_{i=-\infty} a_i \partial^i\quad , \quad a_i \in R \quad
\mbox{{}for all} \;\; i ,
\]
that are added in an obvious way and multiplied according to
\[
\partial^j\circ a \partial^i
=\sum_{k=0}^\infty {j \choose k}\partial^k(a)\partial^{i+j-k}.
\]
Each operator $P=\sum
p_j \partial^j$ decomposes as $P=P_+ + P_-$ with $P_+ =
\sum\limits_{j\geq 0} p_j \partial^j$ its differential operator part
and $P_- = \sum\limits_{j<0} p_j \partial^j$ its integral operator
part. We denote by $Res_\partial P=p_{-1}$ the {\it residue } of $P$.
On $R[\partial , \partial^{-1})$ we have an anti-algebra
morphism called {\it taking the adjoint}. The adjoint of $P=\sum p_i
\partial^i$ is given by
\[
P^* = \sum_i (-\partial)^i p_i .
\]
Further one has a set of derivations $\{ \partial_n \mid n\geq 1\}$ of
$R$ that commute with $\partial$. The equations of the hierarchy can be
{}formulated in a compact way
in a set of relations {}for a so-called {\it Lax operator} in $R[\partial
, \partial^{-1})$, i.e. an operator of the {}form
\begin{equation}
  \label{1.1.1}
  L=\partial + \sum_{j<0} \ell_j \partial^j\quad , \ell_j \in R\quad
  \mbox{{}for all}\;\; j<0.
\end{equation}
These equations are
\begin{equation}
  \label{1.1.2}
  \partial_n (L) = \sum_{j<0} \partial_n (\ell_j) \partial^j =
  [(L^n)_+ , L],\quad n\geq 1.
\end{equation}
Since this equations {}for $n=1$ boils down to $\partial_1 (\ell_j) =
\partial (\ell_j)$ {}for all $j$, we assume from now on that $\partial =
\partial_1$. Equation (\ref{1.1.2}) has at least the trivial solution
$L=\partial$ and can be seen as the compatibility equation of the
linear system
\begin{equation}
  \label{1.1.3}
  L\psi = z\psi \quad \mbox{and}\quad \partial_n (\psi) = (L^n)_+ (\psi)
\end{equation}
One needs a context in which the actions of (\ref{1.1.3}) make
sense and that allows you to derive (\ref{1.1.2}) from
(\ref{1.1.3}). {}For the trivial solution (\ref{1.1.3}) becomes
\[
\partial \psi = z\psi \quad \mbox{and}\quad \partial_n \psi = z^n \psi
\quad \mbox{{}for all}\;\; n\geq 1.
\]
Hence if one takes $\partial_n = \frac{\partial}{\partial t_n}$ then
the function $\gamma(z) = \exp (\sum\limits_{i\geq 1} t_i z^i)$ is a
solution. The space $M$ of so-called {\it oscillating functions} {}for which we
make sense of (\ref{1.1.3}) can
be seen as a collection of perturbations of this solution. It is
defined as
\[
M=\{ (\sum_{i\leq N} a_i z^i) e^{\sum t_i z^i} \mid a_i \in R,
\;\;\mbox{{}for all}\;\; i\}.
\]
The space $M$ becomes a $R[\partial, \partial^{-1})$-module by the
natural extension of the actions
\[
\begin{array}[h]{lcl}
b\{(\sum_j a_j z^j) e^{\sum t_i z^i}\} &=& (\sum_j ba_j z^j) e^{\sum t_i
  z^i}\\
\partial\{ (\sum_j a_j z^j) e^{\sum t_i z^i}\} &=& (\sum_j \partial(a_j)
z^j + \sum_j a_j z^{j+1} ) e^{\sum t_i z^i} .
\end{array}
\]
It is even a free $R[\partial , \partial^{-1})$-module, since we have
\[
(\sum p_j \partial^j) e^{\sum t_i z^i} = (\sum p_j z^j )e^{\sum t_i
  z^i} .
\]

An element $\psi$ in $M$ is called  an {\it oscillating function of type}
$z^{\ell}$,
if it has the
form
\[
\psi (z) = \{ z^{\ell} + \sum_{j<\ell} \alpha_j z^j\} e^{\sum t_i z^i} .
\]
The fact that $M$ is a free $R[\partial, \partial^{-1})$-module,
permits you to show that each oscillating function of type $z^{\ell}$ that
satisfies (\ref{1.1.3}) gives you a solution of (\ref{1.1.2}). This
function is then called a {\it wavefunction} of the KP-hierarchy.

Segal and Wilson give in \cite{SW} an analytic approach to construct
wavefunctions of the KP-hierarchy. They considered the Hilbert space
\[
H=\{ \sum_{n\in \mathbb{Z}} a_n z^n \mid a_n \in \mathbb{C} ,
\sum_{n\in\mathbb{Z}} \mid a_n\mid^2 < \infty\},
\]
with decomposition $H=H_+ \oplus H_-$, where
\[
H_+ = \{ \sum_{n\geq 0} a_n z^n \in H\} \quad \mbox{and}\quad H_- = \{
\sum_{n<0} a_n z^n \in H\}
\]
and inner product
$<\cdot \mid \cdot >$ given by
\[
< \sum_{n\in \mathbb{Z}} a_n z^n \mid \sum_{m\in \mathbb{Z}} b_m z^m >
  = \sum_{n\in \mathbb{Z}} a_n \overline{b_n} .
\]
To this decomposition is associated the Grassmannian $Gr(H)$ consisting
of all closed subspaces $W$ of $H$ such that the orthogonal projection
$p_+ : W\rightarrow H_+$ is Fredholm and the orthogonal projection
$p_- : W \rightarrow H_-$ is Hilbert-Schmidt. The connected components
of $Gr(H)$ are given by
\[
Gr^{(\ell)} (H) = \left\{
 W\in Gr(H) |\;
p_+:z^{\ell}W\rightarrow  H_+ \;\;\mbox{has index zero}
\right\}.
\]
On each of these components we have a natural action by multiplication
of the group of commuting flows
\[
\Gamma_+ = \{ \exp (\sum_{i\geq 1} t_i z^i )\mid t_i \in \mathbb{C} ,\
\sum \mid t_i \mid (1+\epsilon)^i < \infty \quad \mbox{{}for some}\;\;
\epsilon > 0\}.
\]
Now we take {}for $R$ the ring of meromorphic functions on $\Gamma_+$
and {}for $\partial_n$ the partial derivative w.r.t. $t_n$. Then there
exists {}for each $W$ in $Gr^{(-\ell)} (H)$ a wavefunction $\psi_W$ of
type $z^{\ell}$ that is defined on a dense open subset of $\Gamma_+$
and that takes values in $W$. Moreover, it is known that the range of
$\psi_W$ spans a dense subspace of $W$. Hence, if we write $\psi_W =
P_W \cdot e^{\sum t_n z^i}$ with $P_W \in R[\partial ,
\partial^{-1})$, then $L_W = P_W \partial P_W^{-1}$ is a solution of
the KP-hierarchy. Each component of $Gr(H)$ generates in this way the
same set of solutions of the KP-hierarchy, so it would suffice, as is done
in \cite{SW}, to consider only $Gr^{(0)} (H)$. However, it is more
convenient here to consider all components.

A subsystem of the KP-hierarchy consists of all solutions $L$ that are
the $k$-th root of a differential operator. This gives you solutions
of the KP-hierarchy that do not depend on the $\{ t_{kn},\;
\mbox{with}\;\; n\geq 1\}$. Those operators satisfy the condition $L^k
= (L^k)_+$. The set of equations corresponding to this condition is
called the $k$-th Gelfand-Dickey hierarchy. Now it has been shown
that, among the solutions coming from the Segal-Wilson Grassmannian,
the ones that satisfy the $k$-th Gelfand-Dickey hierarchy are exactly
characterized by $z^k W \subset W$. In the next section we consider a
generalization of this condition.

\section{An extension of the condition $z^k W\subset W$}

In this section we consider, {}for each $k$ and $m$  in
$\mathbb{N}=\{ 0,1,2,\ldots\}$, $k\ne 0$
subspaces $W$ in $Gr(H)$ that possess the {\bf  $m$-Vector
$k$-Constrained} ($mVkC$){\bf -condition}:
\begin{equation}
\label{1.2.1}
There\ is\ a\ subspace\ W'\ of\ W\ of\ codimension\ m\ such\ that\ z^k(W')
\subset W.
\end{equation}
This is a natural generalization of the condition that describes
inside $Gr(H)$ the solutions of the $k$-th Gelfand-Dickey
hierarchy. We will show here in a geometric way how you can associate
to each $W$, satisfying the $mVkC$-condition, $2m$ functions $\{ q_j
\mid 1\leq j\leq m\}$ and $\{ r_j \mid 1 \leq j \leq m\}$ {}for which
the following equations hold:
\begin{equation}
  \label{1.2.2}
\partial_n (q_j) = (L^n_W)_+ (q_j) \quad \mbox{{}for all}\;\;n\geq 1,
\end{equation}
\begin{equation}
  \label{1.2.3}
  \partial_n (r_j) = -(L^n_W)^*_+ (r_j) \quad \mbox{{}for all} \;\;n\geq 1.
\end{equation}
Here $A^*$ denotes the adjoint of $A$ in $R[\partial,
\partial^{-1})$. Moreover $L_W$ satisfies
\begin{equation}
  \label{1.2.4}
  L^k_W = (L^k_W)_+ + \sum^m_{j=1} q_j \partial^{-1} r_j.
\end{equation}
At the same time we will give links with the paper of Zhang \cite{Z}.

Take any $W$ in $Gr^{(-\ell)}$ that satisfies the $mVkC$-condition. It
is no restriction to assume that the $m$ occurring in (\ref{1.2.1}) is
optimal, i.e. there is an orthonormal basis $\{ u_1, \ldots , u_m\}$
of the orthocomplement of $W'$ in $W$ such that
\[
(\mbox{Span}\{ z^k u_1 , \ldots , z^k u_m\})\cap W = \{0\}.
\]
Since multiplication with $z$ is unitary, the vectors $\{ z^k (u_1) ,
\ldots , z^k (u_m)\}$ are an orthonormal basis of the orthocomplement
of $W$ in $z^k W + W$. To the space $W$ we associate the subspaces
\[
W_j = W \oplus \mathbb{C} z^k u_j , 1\leq j \leq m.
\]
Clearly the $W_j$ all belong to $Gr^{(-\ell +1)}$ and hence, they have
wavefunctions $\psi_{W_j}$ of type $z^{\ell -1}$ i.e.
\begin{equation}
  \label{1.2.5}
  \psi_{W_j} = \psi_{W_j} (t,z) = \{ z^{\ell - 1} + \sum_{s\geq 1}
  a_{js} (t) z^{\ell -1 -s} \} e^{\sum t_i z^i} .
\end{equation}
Recall that $\psi_{W_j} (t,z)$ is well-defined {}for all $t$ belonging
to the open dense subset
\[
\Gamma^{W_j}_+ = \{ \gamma (z) = \exp (\sum t_i z^i)\in
\Gamma_+ | \gamma^{-1} W_j \;\;\mbox{is transverse to}\;\; z^{\ell -
  1} H_+ \}.
\]
On $\Gamma^{W_j}_+$ we consider the function
\begin{equation}
  \label{1.2.6}
  s_j (t) = < \psi_{W_j} (t,z) \mid z^k u_j >.
\end{equation}
Since the vectors $\{ \psi_{W_j} (t,z) \mid t\in \Gamma^{W_j}_{+}\}$
are lying dense in $W_j$ and $m$ was assumed to be optimal, the
functions $\{s_j\}$ do not vanish.
Hence, on a dense open subset of $\Gamma_+$, there is defined the
function
\begin{equation}
  \label{1.2.7}
  \varphi_j = \frac{1}{s_j} \psi_{W_j} : = r_j \psi_{W_j} .
\end{equation}
It takes values in $W_j$ and has moreover the following useful
property
\begin{equation}
  \label{1.2.8}
  \varphi_j (t) - z^k u_j \in W ,
\end{equation}
{}for all $t$ in a dense open subset of $\Gamma_+$. This property is a
consequence of the facts that $\varphi_j (t) - z^k u_j$ is by
construction orthogonal to $z^k u_j$ and that $W$ is the
orthocomplement of $\mathbb{C} z^k u_j$ inside $W_j$. In \cite{Z},
similar functions $\{ \varphi_j\}$ are introduced, only not using
the geometry, but as solutions of a certain system of differential
equations. In particular, we can dispose of the condition (a) in the
Proposition  of \cite{Z}. Thus we have obtained  $m$  functions $\{r_j\}$.

To define the $\{ q_j\}$ we consider
\begin{equation}
  \label{1.2.9}
  z^k \psi_W - (L^k_W)_+ (\psi_W) = (L^k_W)_- (\psi_W) = \{ \sum_{s
    \geq 0} b_s (t) z^{\ell - 1 - s} \} e^{\sum t_i z^i}.
\end{equation}

{}For each $j, 1\leq j \leq m$, we have a function $q_j$ on
$\Gamma_+^{W_j}$.
\[
\begin{array}[h]{lcl}
q_j (t) &=& <z^k \psi_W (t,z)-(L^k_W)_+ \psi_W (t,z)\mid z^k u^j>\\
        &=& <z^k \psi_W (t,z) \mid z^k u_j>\\
        &=& <\psi_W (t,z) \mid u_j>.
\end{array}
\]
Because $m$ is optimal, the functions $\{ q_j\}$ are non-zero on an
open dense subset of $\Gamma_+$. Since $u_j$ does not depend on $t$
and since $\frac{\partial}{\partial t_n} \psi_W = (L^n_W)_+ (\psi_W)$,
we get directly {}for $q_j$
\begin{equation}
  \label{1.2.10}
  \begin{array}[h]{lcl}
\frac{\partial q_j}{\partial t_n} &=& <\frac{\partial}{\partial t_n}
(\psi_W) (t,z) \mid u_j > = <(L^n_W)_+ (\psi_W (t,z))\mid u_j>\\
&=& (L^n_W)_+ (<\psi_W \mid u_j >) = (L^n_W)_+(q_j).
  \end{array}
\end{equation}
Thus the equations (\ref{1.2.2}) {}for the derivatives of the $\{q_j\}$
are clear. Those {}for the $\{ r_j\}$ require more work.

First we derive an expression {}for $(L^k_W)_- (\psi_W)$. Thereto we
consider
\begin{equation}
  \label{1.2.11}
  \Phi (t) = z^k \psi_W - (L^k_W)_+ (\psi_W) - \sum_{j=1}^m q_j \varphi_j .
\end{equation}
Since $\varphi_j$ takes values in $W_j, the function (L^k_W)_+
(\psi_W)$ does so in the
space $W$ and $z^k \psi_W$ in $z^k W$. Hence we have that $\Phi (t)$ belongs
to $W + z^k W$ {}for all relevant $t$. By construction we have that {}for
all $j, 1\leq j \leq m, \Phi (t)$ is orthogonal to $z^k u_j$, hence
$\Phi (t)$ even belongs to $W$. From the {}form of the $\varphi_j$, we
see that on an open dense set of $\Gamma_+$ one has
\[
\Phi (t) = \{ \sum_{s\geq 0} c_s z^{\ell - 1 - s}\} e^{\sum t_i z^i} .
\]
By construction, there holds
\[
W\cap (z^{\ell} H_+)^{\perp} \gamma(z) = \{ 0 \} ,
\]
so that we arrive at
\begin{equation}
  \label{1.2.12}
  z^k \psi_W - (L^k_W)_+ (\psi_W) = \sum^m_{j=1} q_j \varphi_j .
\end{equation}
This equation is part of the system of differential equations {}for the
$\varphi_j$ as used in $[Z]$. Recall that $\varphi_j$ has the {}form
\[
\varphi_j = \{ r_j z^{\ell - 1} + \;\;\mbox{lower order terms in}\;\;
z\} e^{\sum t_i z^i}.
\]
Hence,
\[
\frac{\partial \varphi_j}{\partial x} = \frac{\partial
  \varphi_j}{\partial t_1} = \{ r_j z^{\ell} + \;\;\mbox{lower order
  terms}\; \} e^{\sum t_i z^i}.
\]
On the other hand we know that $\varphi_j (t) - z^k u_j$ belongs to
$W$ {}for all $t$. Thus also $\frac{\partial \varphi_j}{\partial x} (t)$
belongs to $W$. In $W$ we have that
\[
\frac{\partial \varphi_j}{\partial x} - r_j \psi_W = \{ \sum_{s\geq 0}
\alpha_s z^{\ell - 1 - s} \} e^{\sum t_i z^i} \in (z^{\ell}
  H_+)^{\perp} \gamma
\]
and this has to be zero. By definition we have $\varphi_j = r_j
\psi_{W_j}$ and differentiation w.r.t. $x$ gives
\begin{equation}
  \label{1.2.13}
  \psi_W = \frac{1}{r_j} \partial (r_j \psi_{W_j}) = (r^{-1}_j
  \partial r_j)(\psi_{W_j}).
\end{equation}
Consequently, we have {}for $\phi_j$
\[
\varphi_j = r_j \psi_{W_j} = r_j (r^{-1}_j \partial^{-1} r_j) \psi_W =
\partial^{-1} r_j \psi_{W_j}.
\]
Now we substitute this in equation (\ref{1.2.12}) and obtain
\begin{equation}
  \label{1.2.14}
  (L^k_W)_- (\psi_W) = \{ \sum^m_{j=1} q_j \partial^{-1} r_j \} \psi_W.
\end{equation}
Since the pseudodifferential operators act freely on wavefunctions, we
see that $L_W$ and the functions $\{q_j\}$ and $\{ r_j\}$ are exactly
connected by equation (\ref{1.2.4})
\[
(L^k_{W})_- = \sum^m_{j=1} q_j \partial^{-1} r_j .
\]
What remains to be shown, is the differential equation (\ref{1.2.3}) {}for the
$r_j$. As $\varphi_j (t) - z^k u_j$ belongs to $W$, it follows that
{}for all $n\geq 1 , \frac{\partial \varphi_j}{\partial t_n} (t)$ lies in
$W$. Recall that
\[
\varphi_j = \{ r_j z^{\ell -1} + \;\;\mbox{lower order terms
  in}\;\;z\} e^{\sum t_i z^i}.
\]
Then we have
\[
\begin{array}[h]{lcl}
{\displaystyle {\partial \varphi_j}\over\displaystyle{\partial t_n}} &=&
\{r_j z^{n+\ell -1} + \;\;\mbox{lower order terms}\} e^{\sum t_i
  z^i}\\
&=& \{ r_j \partial^{n-1} \} \psi_W + \{\sum_{s\geq 0} \alpha_s
z^{n-1+\ell -s} \} e^{\sum t_i z^i}\\
&=& A_{nj} (\psi_W) + \{ \sum_{s\geq 0} \beta_s z^{\ell -1-s}\}
e^{\sum t_i z^i},
\end{array}
\]
with $A_{nj}$ a uniquely determined differential operator in
$\partial$ of order $n-1$ and with leading coefficient $r_j$. Since
both $ \frac{\partial \varphi_j}{\partial t_n}$ as $A_{nj} (\psi_W)$
are lying in $W$, we get
\[
 \frac{\partial \varphi_j}{\partial t_n} - A_{nj} (\psi_W) = 0 = W\cap
 (z^{\ell} H_+)^{\perp} \gamma(z) .
\]
On the other hand we know that $\varphi_j = \partial^{-1} r_j \psi_W$
and this leads to
\begin{equation}
  \label{1.2.15}
  A_{nj} (\psi_W) = \partial^{-1}  \frac{\partial r_j}{\partial
    t_n} \psi_W + \partial^{-1} r_j (L^n_W)_+ (\psi_W) .
\end{equation}
This gives you an expression {}for $A_{nj}$ in $L_W$ and $r_j$
\[
A_{nj} = \partial^{-1} ( \frac{\partial r_j}{\partial t_n} + r_j
(L^n_W)_+ ).
\]
By taking the residue in $\partial$ of the operators in this equation,
we see that
\[
\mbox{Res}_{\partial} (A_{nj})=0 =  \frac{\partial r_j}{\partial t_n}
+ \mbox{Res}_{\partial} (\partial^{-1} r_j (L^n_W)_+ )= \frac{\partial
  r_j}{\partial t_n} + (L^n_W)^*_+ (r_j) .
\]
The last equality is a direct consequence of the following property of
residues of pseudodifferential operators.

\begin{lemma}
  \label{L2.1}
In the ring $R[\partial , \partial^{-1})$ of pseudodifferential operators
with coefficients in $R$, we have {}for each $f$ in $R$ and
$P=\sum_{j\leq N} p_j \partial^j$ in $R[\partial , \partial^{-1})$
\[
\mbox{Res}_{\partial} (\partial^{-1} f P) = (P^*)_+ (f) ,
\]
where $(P^*)_+ = \sum\limits_{0\leq j\leq N} (-\partial)^j p_j$ is the
differential operator part of the adjoint of $P$.
\end{lemma}

\noindent {\bf Proof.}
First we recall that $\mbox{Res}_{\partial}$ behaves as follows w.r.t. to
taking the adjoint $P^* = \sum\limits_{j\leq N} (-\partial)^j p_j$ of
$P$
\[
\mbox{Res}_{\partial} (P^*) = -\mbox{Res}_{\partial} P .
\]
This is easily reduced to operators of the {}form $a\partial^n,
n\in\mathbb{Z}$. Next one notices that it suffices to prove the
equality in the lemma {}for differential operators. The left hand side
{}for such a $P$ trans{}forms as
\[
\mbox{Res}_{\partial} (\partial^{-1} f P) = -\mbox{Res}_{\partial} (P^*
f(-\partial)^{-1}) = \mbox{Res}_{\partial} (P^* f\partial^{-1}) .
\]
As $P^* f$ is a differential operator with constant term $P^* (f)$,
this gives the proof of the lemma.
\hfill{$\Box$}
\\

\noindent So we have shown that each $r_j$ satisfies the equation
(\ref{1.2.3}):
\[
 \frac{\partial r_j}{\partial t_n} = -(L^n_W)_+ (r_j).
\]
and we can conclude that $L_W$, the $\{q_j\}$ and the $\{ r_j\}$ {}form
a solution of the $m$ vector $k$-constrained KP-hierarchy.

\section{The main theorem}

In this subsection we will prove the converse of the result from the
{}foregoing subsection and thus come to the main theorem. So we start
with a $W$ in $Gr^{(-\ell)}$ and functions $\{ q_j\}$ and $\{r_j\}$,
all defined on a dense open subset of $\Gamma_+$, such that the equations
(\ref{1.2.2}) , (\ref{1.2.3}) and (\ref{1.2.4}) are satisfied. We will
show that such a $W$ fulfills the $mVkC$-condition from section
3. Recall that there is a unique pseudodifferential operator $P_W$
such that $\psi_W = P_W (e^{\sum t_i z^i})$. It has the {}form
\begin{equation}
\label{4.0}
P_W = \partial^{\ell} + \sum_{j<\ell} p_j \partial^j = \{ 1 +
\sum_{s<0} p_{\ell + s} \partial^s \} \partial^{\ell}.
\end{equation}
It is not difficult to see that the fact that $\psi_W $ is a
wavefunction
is equivalent to $P_W$
satisfying the Sato-Wilson equations
\begin{equation}
  \label{1.3.1}
   \frac{\partial P_W}{\partial t_n} P^{-1}_W = - (P_W
   \partial^n P_W^{-1})_-,
\end{equation}
where $P_-$ denotes  the integral operator part $\sum\limits_{i<0}
p_i \partial^i$ of the element $P=\sum p_j \partial^j$ in $R[\partial
,\partial^{-1})$.
Next we consider {}for each $j, 1\leq j\leq m$, the operators $Q_j$ and
$R_j$ defined by
\begin{equation}
  \label{1.3.2}
  Q_j := q_j \partial q_j^{-1} P_W \quad \mbox{and}\quad R_j =
  r_j^{-1} \partial^{-1} r_j P_W.
\end{equation}
We want to show that the $Q_j$ and the $R_j$ also satisfy the
Sato-Wilson equations. To do so, we need some properties of the ring
$R[\partial , \partial^{-1})$ of pseudodifferential operators with
coefficients from $R$. We resume them in a lemma

\begin{lemma}
  \label{L1.3.1}
If $f$ belongs to $R$ and $Q$ to $R[\partial , \partial^{-1})$, then
the following identities hold
\begin{enumerate}
\item[(a)] $(Q f)_- = Q_- f$,
\item[(b)] $(fQ)_- = fQ_-$,
\item[(c)] Res$_{\partial}(Qf) =$ Res$_{\partial} (fQ) = f$ Res$_\partial (Q)$,
\item[(d)] $(\partial Q)_- = \partial Q_- -$ Res$_{\partial} (Q)$,
\item[(e)] $(Q\partial )_- = Q_- \partial-$ Res$_{\partial} (Q)$,
\item[(f)] $(Q\partial^{-1})_- = Q_- \partial^{-1} +$ Res$_{\partial}
  (Q\partial^{-1}) \partial^{-1}$,
\item[(g)] $(\partial^{-1} Q)_- = \partial^{-1} Q_- + \partial^{-1}$
  Res$_{\partial} (Q^* \partial^{-1})$.
\end{enumerate}
\end{lemma}
Since the proof of this lemma consists of straight{}forward
calculations, we leave this to the reader. Now we can show

\begin{proposition}
  \label{P1.3.2}
The operators $Q_j$ and $R_j, 1\leq j \leq m$, satisfy the Sato-Wilson
equations.
\end{proposition}

\noindent{\bf Proof.}
If we denote $ \frac{\partial}{\partial t_n}$ by $\partial_n$, then we
get {}for $Q_j = q_j \partial q_1^{-1} P_W$ that
\[
\begin{array}[lcl]{lcl}
\partial_n (Q_j)Q^{-1}_j &=&
\partial_n (q_j \partial q^{-1}_j) q_j \partial^{-1}_j q^{-1}_j + q_j
\partial q^{-1}_j \partial_n (P_W) P^{-1}_W q_j \partial^{-1}
q_j^{-1}\\
&=& -q_j \partial q_j^{-1} (L^n_W)_- q_j \partial^{-1} q^{-1}_j +
\partial_n (q_j \partial q^{-1}_j) q_j \partial^{-1} q^{-1}_j.
\end{array}
\]
Now we apply successively the identities from Lemma \ref{L1.3.1} to the
first operator of the right-hand side
\[
\begin{array}[h]{lclc}
q_j \partial q_j^{-1} (L^n_W)_- q_j \partial^{-1} q_j &=&
q_j \partial (q_j^{-1} L^n_W q_j) \partial^{-1} q_j^{-1} &=\\

q_j \partial (q_j)^{-1} L^n_W q_j \partial^{-1})_- q_j^{-1} &-&
q_j \partial \mbox{Res}_{\partial} (q_j^{-1} L^n_W q_j
\partial^{-1})\partial^{-1} q_j^{-1} &=\\

q_j (\partial q_j^{-1} L^n_W q_j \partial^{-1})_- q^{-1}_j &+&
q_j \mbox{Res}_{\partial} (q^{-1}_j L^n_W q_j \partial^{-1}) q^{-1}_j
&-\\

q_j \partial \mbox{Res}_{\partial} (q_j^{-1} L^n_W q_j \partial^{-1})
\partial^{-1} q^{-1}_j &=&
(q_j \partial q_j^{-1} L^n_W q_j \partial^{-1} q_j^{-1})_- &+\\

q^{-1}_j \mbox{Res} (L^n_W q_j \partial^{-1}) &-&
q_j \partial q_j^{-1} \mbox{Res}_{\partial} (L^n_W q_j \partial^{-1})
\partial^{-1} q_j^{-1}.
\end{array}
\]
By applying Lemma \ref{1.2.1} to these last two residues we get
\[
(q_j \partial q_j^{-1} L^n_W q_j \partial^{-1} q_j^{-1})_- + (L^n_W)_+
  (q_j) q_j^{-1} - q_j \partial q_j^{-1} (L^n_W)_+ (q_j) \partial^{-1}
q_j^{-1}.
\]
On the other hand
\[
\partial_n (q_j \partial q_j^{-1}) q_j \partial^{-1} q_j^{-1} =
\partial_n (q_j) q^{-1}_j - q_j \partial q^{-2}_j \partial_n (q_j) q_j
\partial^{-1} q_j^{-1}.
\]
Thus we see that, if $\partial_n (q_j) = (L^n_W)_+ (q_j)$, the
operator $Q_j$ satisfies the Sato-Wilson equation
\begin{equation}
  \label{1.3.3}
  \partial_n (Q_j) Q_j^{-1} = -(Q_j \partial^n Q_j^{-1})_- .
\end{equation}
{}For $R_j$, we proceed in a similar fashion
\[
\begin{array}[h]{l}
\partial_n (R_j) R^{-1}_j = -r_j^{-1} \partial^{-1} r_j (L^n_W)_- r_j
\partial r_j + \partial_n (r_j^{-1} \partial^{-1} r_j) r_j^{-1}
\partial r_j\\
= -r_j^{-1} \partial^{-1} (r_j L^n_W r^{-1}_j)_- \partial r_j +
-\partial_n (r_j) r_j^{-1} + r^{-1}_j \partial^{-1} (\partial_n (r_j)
r_j^{-1})\partial r_j.
\end{array}
\]
Now we successively apply Lemma \ref{L1.3.1} (g) and (c) and (\ref{1.3.1}) to
the first term of the right hand side of this equation
\[
\begin{array}[h]{clclcl}
 -& r_j^{-1}\partial^{-1} (r_j L^n_W r_j^{-1})_- \partial r_j
&=& -r_j^{-1} \{(\partial^{-1} r_j L^n_W r_j^{-1} )_-
&-& \partial^{-1} \mbox{Res}(r_j^{-1} (L^n_W)^* r_j \partial^{-1})\}
    \partial r_j\\
 =& -r_j^{-1} (\partial^{-1} r_j L^n_W r_j^{-1})_- \partial r_j
&+& r_j^{-1} \partial^{-1} r_j^{-1} (L^n_W)^* (r_j) \partial r_j
& & \\
 =& -r_j^{-1} \{ (\partial^{-1} r_j L^n_W r_j^{-1} \partial)_-
&+& \mbox{Res}_{\partial} (\partial^{-1} r_j L^n_W r_j^{-1}) \} r_j
&+& r_j^{-1} \partial^{-1} r_j^{-1} (L^n_W)^* (r_j) \partial r_j\\
 =& -(r_j^{-1} \partial^{-1} r_j L^n_W \partial r_j)_-
&-& r_j^{-1} (L^n_W)^*_+ (r_j)
&+& r_j^{-1} \partial^{-1} r_j^{-1} (L^n_W)^* (r_j) \partial r_j.
\end{array}
\]
Since $\partial_n (t_j) = -(L^n_W)^* (r_j)$, we see that the last two
terms cancel $\partial_n (r_j^{-1} \partial r_j) r_j^{-1}
\partial r_j$ and thus we have obtained the Sato-Wilson equation {}for
$R_j$
\begin{equation}
  \label{1.3.5}
  \partial_n (R_j) R_j = -(R_j \partial^n R_j^{-1})_-.
\end{equation}
This concludes the proof of proposition \ref{P1.3.2}.
\hfill{$\Box$}
\\

\noindent This proposition has some important consequences. Since the $\{
r_j\}$
and the $\{ q_j\}$ are non-zero on a dense open subset of
$\Gamma_+$, we define on such a subset of $\Gamma_+$ oscillating
functions $\psi_{Q_j}$ and $\psi_{R_j}$ of type $z^{\ell +1}$
  resp. $z^{\ell -1}$ by
  \begin{equation}
    \label{1.3.6}
    \psi_{Q_j} = q_j \partial q^{-1}_j \cdot \psi_W \quad
    \mbox{and}\quad \psi_{R_j} = r_j^{-1} \partial^{-1} r_j \cdot \psi_W.
  \end{equation}
Consider the following subspaces in $Gr(H)$
\[
W_{Q_j} = \overline{\mbox{Span} \{ \psi_{Q_j} (t,z)\}} \quad
\mbox{and}\quad W_{R_j} = \overline{\mbox{Span}\{\psi_{R_j} (t,z)\}} .
\]
Then we can conclude from proposition \ref{P1.3.2}
\begin{corollary}
  \label{C4.3}
The functions $\psi_{Q_j}$ and $\psi_{R_j}$ are the wavefunctions of
the planes $W_{Q_j}$ and $W_{R_j}$. Moreover we have the following
codimension 1 inclusions:
\[
W_{Q_j} \subset W \quad \mbox{and}\quad W\subset W_{R_j}
\]
\end{corollary}

\noindent {\bf Proof.}
{}From the Sato-Wilson equations one deduces directly that {}for all
$n\geq 1$,
\[
\partial_n \psi_{Q_j} = (Q_j \partial^n Q_j^{-1})_+ \psi_{Q_j}
\quad \mbox{and}\quad
\partial_n \psi_{R_j} = (R_j \partial^n R^{-1}_j)_+ \psi_{R_j}.
\]
This shows the first part of the claim. The inclusions between the
different spaces follows from the relations
\[
\psi_{Q_j} = (q_j \partial q^{-1}_j)(\psi_W)
\quad \mbox{and} \quad
\psi_W = (r_j \partial r_j^{-1}) \psi_{R_j}
\]
and the fact that the values of a wavefunction corresponding to an
element of $Gr(H)$ are lying dense in that space. Since {}for a suitable
$\gamma$ in $\Gamma_+$ the orthogonal projections of $\gamma^{-1}
W_{R_j}$ on $z^{\ell} H_+$ resp. $\gamma^{-1} W$ on $z^{\ell + 1} H_+$
have a one dimensional kernel, one obtains the codimension one
result. This concludes the proof of the corollary.
\hfill{$\Box$}
\\
\noindent
Now we can {}formulate the main results of this paper.

\begin{theorem}
  \label{T1.3.4}
Let $W$ be a plane in $Gr(H)$ and let $L_W$ be the corresponding
solution of the KP-hierarchy. Then {}for $m, k\in\mathbb{N}$,
$k\ne 0$, the following 2 conditions are
equivalent
\begin{enumerate}
\item[(a)] The space $W$ satisfies the $mVkC$-condition.
\item[(b)] There exist functions $\{q_j \mid 1\leq j \leq m\}$ and $\{
  r_j \mid 1\leq j\leq m\}$ defined on an open dense subset of
  $\Gamma_+$ such that the following conditions are fulfilled:
  \begin{enumerate}
  \item[(i)] $\partial_n (q_j) = (L_W^n )_+ (q_j)$
           \quad  {}for all $n\geq 1$,
  \item[(ii)] $\partial_n (r_j) = -(L_W^n)^*_+
              (r_j)$\quad  {}for all $n\geq 1$,
  \item[(iii)] $L_W^k = (L_W^k)_+ + \sum\limits^m_{j=1} q_j
    \partial^{-1} r_j$.
  \end{enumerate}
\end{enumerate}
\end{theorem}

\noindent {\bf Proof.}
In section 2 it has been shown that (a) implies (b). So we assume from
now on (b). The relation (b) (iii) leads to
\[
\begin{array}[h]{lcl}
L_W^k (\psi_W) &=& z^k \psi_W\\
               &=& (L^k_W)_+ (\psi_W) + \sum\limits^m_{j=1} q_j \partial^{-1}
                   r_j \psi_W\\
               &=& (L^k_W)_+ (\psi_W) + \sum\limits_{j\atop r_j\ne 0} q_j r_j
r_j^{-1}
                    \partial^{-1} r_j \psi_W\\
               &=& (L^k_W)_+ (\psi_W) + \sum\limits_{j\atop r_j \neq 0}
                      q_j r_j \psi_{R_j}.
\end{array}
\]
Thus we see with the usual density argument that
\[
z^k W\subset W + \sum_j W_{R_j} = \sum_j W_{R_j} = \tilde{W} .
\]
Since each $W$ has codimension one in $W_{R_j}$, we see that the
codimension of $W$ in $\tilde{W}$ is $\leq m$. Let $W_1$ be the
orthocomplement of $W$ in $\tilde{W}$ and $p_1 : H\rightarrow W_1$ the
orthogonal projection on $W_1$. Inside $W$ we consider
\[
W^1 = \{ w\in W\mid p_1 (z^k w) = 0\}.
\]
Since $\dim (W_1)\leq m$, we see that $W^1$ is a subspace of $W$ of
codimension $\leq m$ and by construction $z^k W^1 \subset W$. This
completes the proof of the theorem.
\hfill{$\Box$}

\section{General rational reductions of the KP hierarchy}

We are now going to connect the vector constrained KP hierarchy to
reductions of the KP hierarchy introduced by Krichever \cite{K}.
For that purpose we assume that $W$ is a plane
in $Gr(H)$ that satisfies the $mVkC$-condition, where we choose $m$ to
be as minimal as is  possible for that plane. Let $L_W=P_W\partial P_W^{-1}$,
with
$P_W$ of the form (\ref{4.0}), be the corresponding solution of the KP
hierarchy and let $W^1\subset W$ be the subspace of codimension $M$
such that $W_1=z^kW^1\subset W$. Notice first that  $W_1$ is  a
subspace of $W$ and $z^kW$ of codimension $k+m$ and $m$, respectively.
Hence there exist
differential operators $L_1$ and $L_2$ of order $k+m$ and $m$,
respectively, such that
\begin{equation}
\label{5.1}
L_1\psi_W=\psi_{W_1},\quad L_2z^k\psi_W=\psi_{W_1}
\end{equation}
and that $\psi_{W_1}$ is again a wavefunction. From (\ref{5.1})
one immediately deduces that
\begin{equation}
\label{5.2}
L_W^k=L_2^{-1}L_1.
\end{equation}
We first  prove the following lemma.
\begin{lemma}
  \label{L5.1}
Let $L=P\partial^kP^{-1}$ be a pseudodifferential operator of order
$k$ and let $L_1$ and $L_2$ be differential operators of order $k+m$
and $m$, respectively, such that $L=L_2^{-1}L_1$. Then one has the
following identities:
\[
L_1(L_2^{-1}L_1)^{i/k}=(L_1L_2^{-1})^{i/k}L_1,\quad
L_2(L_2^{-1}L_1)^{i/k}=(L_1L_2^{-1})^{i/k}L_2.
\]
\end{lemma}
\noindent {\bf Proof.}
 Since $L_1P=L_2P\partial^k$, one can find a pseudodifferential
operator $Q$ of the same order as $P$ such that
$L_1=Q\partial^{k+m}P^{-1}$,
$L_2=Q\partial^mP^{-1}$ and thus
$L_1L_2^{-1}=Q\partial^kQ^{-1}$. Since also
$L_2^{-1}L_1=P\partial^kP^{-1}$, one finds that their $k$-th roots
satisfy
\[
(L_2^{-1}L_1)^{1/k}=P\partial P^{-1},\quad
(L_1L_2^{-1})^{1/k}=Q\partial Q^{-1}.
\]
Using this, one easily verifies the identities of the Lemma.
\hfill{$\Box$}

\
\\

\noindent
Since both $\psi_W$ and $\psi_{W_1}$ are wavefunctions that are
connected by equations (\ref{5.1}),
we find, using (\ref{5.2}) and Lemma \ref{L5.1},  that
\begin{equation}
\label{5.3}
L_W=(L_2^{-1}L_1)^{1/k}\quad\text{and}\quad
L_{W_1}=L_1(L_2^{-1}L_1)^{1/k}L_1^{-1}=(L_1L_2^{-1})^{1/k}.
\end{equation}
Hence
\[
\partial_i\psi_{W_1}=((L_1L_2^{-1})^{i/k})_+\psi_{W_1}
=((L_1L_2^{-1})^{i/k})_+L_1\psi_{W}
\]
and on the other hand is also equal to
\[
\partial_i(L_1\psi_{W})
=\partial_i(L_1)\psi_{W}+L_1((L_2^{-1}L_1)^{i/k})_+\psi_{W}.
\]
{}From which one deduces that
\begin{equation}
\label{5.4}
\partial_iL_1=((L_1L_2^{-1})^{i/k})_+L_1-L_1((L_2^{-1}L_1)^{i/k})_+.
\end{equation}
In a similar way one obtains from the other identity of (\ref{5.1})
that
\begin{equation}
\label{5.5}
\partial_iL_2=((L_1L_2^{-1})^{i/k})_+L_2-L_2((L_2^{-1}L_1)^{i/k})_+.
\end{equation}
Notice that in this way we have exactly obtained
Krichever's general rational reductions of the KP hierarchy \cite{K}.
Krichever considers KP pseudodifferential
operators $L$ of the form (\ref{1.1.1}),
such that $L^k=L_2^{-1}L_1$, where $L_1$ and $L_2$
are coprime differential operators  of order $k+m$ and $m$,
respectively. It can be shown that the equations
(\ref{5.4}) and (\ref{5.5}) for $L_1$ and $L_2$ are equivalent to the
KP Lax equations for $L$.
It is not difficult to see that our operators must be coprime, since
we have chosen our $m$ to be minimal.
We will now prove that the converse also holds, i.e, that the
following theorem holds.
\begin{theorem}
  \label{T5.1}
Let $W$ be a plane in $Gr(H)$ and let $L_W$ be the corresponding
solution of the KP-hierarchy. Then {}for $m, k\in\mathbb{N}$,
$k\ne 0$, the following 2 conditions are
equivalent
\begin{enumerate}
\item[(a)] The space $W$ satisfies the $mVkC$-condition, with $m$ as
minimal as possible.
\item[(b)] There exist coprime differential operators $L_1$ and $L_2$
of order $k+m$ and $m$, respectively,
such that the following conditions are fulfilled:
  \begin{enumerate}
  \item[(i)] $L_W^k=L_2^{-1}L_1$
  \item[(ii)]
$\partial_iL_1=((L_1L_2^{-1})^{i/k})_+L_1-L_1((L_2^{-1}L_1)^{i/k})_+$,
  \item[(iii)]
$\partial_iL_2=((L_1L_2^{-1})^{i/k})_+L_2-L_2((L_2^{-1}L_1)^{i/k})_+$.
  \end{enumerate}
\end{enumerate}
\end{theorem}
\noindent {\bf Proof.}
We have already shown that (a) implies (b). So we assume from now on
(b). Let $\psi_1$ be the oscillating function $L_1\psi_W$,
then by using Lemma \ref{L5.1}:
\[
(L_1L_2^{-1})^{1/k}\psi_{1}=(L_1L_2^{-1})^{1/k}L_1\psi_{W}
=L_1(L_2^{-1}L_1)^{1/k}\psi_{W}=zL_1\psi_{W}=z\psi_{1}.
\]
Now consider
\[
\begin{array}[h]{lcl}
\partial_i\psi_{1}&=&\partial_i(L_1)\psi_{W}+L_1\partial_i\psi_{W}\\
&=&(((L_1L_2^{-1})^{i/k})_+L_1-L_1((L_2^{-1}L_1)^{i/k})_+
+L_1((L_2^{-1}L_1)^{i/k})_+)\psi_{W}\\
&=&((L_1L_2^{-1})^{i/k})_+L_1\psi_{W}\\
&=&((L_1L_2^{-1})^{i/k})_+\psi_{1}.
\end{array}
\]
Hence $\psi_{1}$ is again a wavefunction of the KP hierarchy. If we
let $W_1$ be the closure of the span of the $\psi_1(t,z)$ then
$\psi_{W_1}=\psi_1$.
Since
$z^k\psi_{W}$ is also a wavefunction,
\[
L_2z^k\psi_{W}=\psi_{W_1}.
\]
Thus we see with the usual density argument that
\begin{equation}
\label{5.6}
\begin{array}[h]{lcl}
W_1&\subset &z^kW\quad\text{of codimension }m\\
W_1&\subset &W\quad\text{of codimension }k+m
\end{array}
\end{equation}
Hence $W^1=z^{-k}W_1$ is a subset of $W$ of codimension $m$ such that
$z^kW^1\subset W$. Since our differential operators are coprime , one
cannot find lower order operators $M_1$ and $M_2$ such that
$L_W=M_2^{-1}M_1$. Hence there is no smaller subspace $W_1$ and no
smaller $m$ such that (\ref{5.6}) is satisfied.
\hfill{$\Box$}
\
\\

\noindent
As a consequence of this, we obtain that in the  Segal-Wilson setting,
the vector constrained KP hierarchy and Krichever's general rational
reduction define the same reduction of the KP hierarchy.

\
\\

\noindent
{\Large \bf  Acknowledgements.} We would like to thank Igor Krichever,
for sending us an early preprint version of his paper \cite{K}, and
especially Henrik Aratyn, for sending us \cite{A0}. In this communication he
presents his proof, that the vector constrained KP and Krichever's general
rational reduction's of KP describe the same hierarchies. His proof is
based on kernels of differential operators and properties of Wronskians
and is quite different from the proof given in this paper.

\end{document}